\documentclass[sigplan,9pt,screen]{acmart}

\usepackage{diagbox}
\usepackage{breqn}
\usepackage{filecontents}
\usepackage{algorithm}
\usepackage{algpseudocode}
\usepackage{multirow}
\usepackage{makecell}
\usepackage{fancyhdr}
\usepackage{listings}
\usepackage{balance}
\usepackage{amsmath,amsfonts}
\usepackage{graphicx}
\usepackage{epstopdf}
\usepackage{textcomp}
\usepackage{xcolor}
\usepackage{subcaption} 
\usepackage{float}
\usepackage{booktabs}
\usepackage{algorithm}
\usepackage{algpseudocode}
\usepackage{setspace}
\usepackage{svg} 
\usepackage{etoolbox}
\usepackage{url}
\usepackage{newtxtext,newtxmath}
\usepackage{tikz}
\usepackage{tabularx}
\usepackage{booktabs}
\usepackage{multirow}
\usepackage{makecell}
\usepackage{array}
\usepackage{cancel}
\usepackage{colortbl}
\usepackage{algpseudocode}
\usepackage{graphicx}
\usepackage[font=small,labelfont=bf,textfont=bf]{caption}
\usepackage{fancyhdr}

\newrobustcmd*\blacka[1]{\tikz[baseline=(char.base)]{
            \node[shape=circle,draw,inner sep=1pt,fill,text=white,minimum size=1em] (char) {\textsf{\small a}};}}

\newrobustcmd*\blackb{%
  \tikz[baseline=(char.base)]{
    \node[shape=circle,draw,fill,text=white,
          minimum size=1em, inner sep=0pt,
          text height=1ex, text depth=0ex,
          font=\sffamily\small] (char) {b};}}

\newrobustcmd*\blackc[1]{\tikz[baseline=(char.base)]{
            \node[shape=circle,draw,inner sep=1pt,fill,text=white,minimum size=1em] (char) {\textsf{\small c}};}}

\newrobustcmd*\blackd{%
  \tikz[baseline=(char.base)]{
    \node[shape=circle,draw,fill,text=white,
          minimum size=1em, inner sep=0pt,
          text height=1ex, text depth=0ex,
          font=\sffamily\small] (char) {d};}}

\setcopyright{none}
\begin{document}
\newcommand{\zhenote}[1]{\textbf{\textcolor{blue}{Zhe: #1}}}
\newcommand{\guan}[1]{\textbf{\textcolor{red}{Guan: #1}}}

\title{From Characterization to Microarchitecture: Designing an Elegant and Reliable BFP-Based NPU}

\author{Jie Zhang$^{\dagger}$$^1$, Jiapeng Guan$^\S${$^1$}, Hao Zhou$^\S$, Xiaomeng Han$^{\dagger}$, Tinglue Wang$^{\dagger}$, Ran Wei$^{\P}$ and Zhe Jiang{$^\dagger$}$^*$}
\thanks{$^1$Both authors contributed equally.}
\thanks{$^*$Corresponding author:
Zhe Jiang. Email: zhejiang.uk@gmail.com.
}

\affiliation{%
  \institution{$^{\dagger}$Southeast University, China \quad $^\S$Dalian University of Technology, China \quad $^{\P}$Lancaster University, UK}
  \country{}
}

\renewcommand{\shortauthors}{Jie Zhang et al.}


\begin{abstract}
Block Floating-Point (BFP) is emerging as an attractive data format for edge Neural Processing Units (NPUs), combining wide dynamic range with high hardware efficiency. 
However, its behavior under hardware faults and suitability for safety-critical deployments remain underexplored.
Here, we present the first in-depth empirical reliability study of BFP-based NPUs.
Using RTL-level fault injection on NPUs, our bit- and path-level analysis reveals pronounced heterogeneous vulnerabilities and shows conventional end-to-end check becomes ineffective under nonlinear block scaling.
Guided by these insights, we design a fault-tolerant BFP-based NPU microarchitecture that aligns the BFP computational semantics with reliability constraints. The design uses a row/column-wise blocking strategy to decouple the fixed-point mantissa computations from the scalar exponent path, and introduces ultra-lightweight protection mechanisms for each.
Experimental results demonstrate our design achieves near–dual modular redundancy reliability with only 3.55\% geometric mean performance overhead and less than 2\% hardware cost.
\end{abstract}

\maketitle
\section{Introduction}
\label{sc:Introduction}
Block Floating-Point (BFP)~\cite{yeh2022like,zou2024bie,darvish2020pushing} is an emerging format gaining traction in modern edge Neural Processing Units (NPUs)~\cite{drumond2018training,haris2024designing,haris2025f}.
In BFP, a ``block'' of elements shares a single common exponent, while each element retains its own mantissa. 
By aligning the exponents within the block, arithmetic operations on the mantissa can be executed entirely using integer units, requiring only a single exponent adjustment per block.
This simplifies the complexity of Multiply–and-Accumulate (MAC) datapaths and Floating-Point (FP) pipelines, enhancing computational throughput and energy efficiency compared to the IEEE-754 FP formats, hence has been widely adopted in many industrial products, e.g., NVIDIA's Blackwell GPUs~\cite{nvidia_mxfp_blog_2025}, AMD's Strix Point NPUs~\cite{amd_blockfp16_blog_2025}, and Tenstorrent AI accelerators~\cite{cavagna2025assessing}.

While performance improvement drives the widespread adoption of BFP, \emph{reliability} is equally important for edge NPUs, especially for the ones deployed in safety-critical scenarios, e.g., autonomous driving~\cite{fu2024drive,li2017understanding}, industrial control~\cite{solowjow2020industrial,topalova2023model}, etc.
This means that the NPU must be resilient to hardware faults and able to detect errors within a narrow Fault Tolerant Time Interval (FTTI), often measured in milliseconds, before escalating into hazards.

\noindent \textbf{Existing work.}
A variety of fault-tolerance techniques have been explored in prior works for edge NPUs using conventional data formats, e.g., INT8/16~\cite{jacob2018quantization,porsia2025resilience}, FP16~\cite{micikevicius2017mixed,fang2023mpgemmfi}, BF16~\cite{kalamkar2019study,fang2023mpgemmfi}.
These works typically start with an offline reliability analysis (e.g., fault injection or analytical modeling~\cite{porsia2025resilience,fang2023mpgemmfi,ping2020sern}) to identify the vulnerable bits, modules, or pipeline stages. 
Based on the analysis, targeted safety mechanisms are then developed and deployed~\cite{bertoa2022fault,ordonez2024enhancing,ahmadilivani2023enhancing}. 
For instance, protecting the systolic array using Algorithm-Based Fault Tolerance (ABFT)~\cite{aliagha2025scissors,safarpour2021algorithm,mummidi2022highly}, redundancy (e.g., hardware Dual-Module Redundancy (DMR)~\cite{bannon2019computer,sheikh2018double}, or software Instruction Redundancy (IR)~\cite{ibrahim2020analyzing,wei2020analyzing,ibrahim2020soft}), and safeguarding registers or memory cells with ECC or parity~\cite{lee2022value,park2025pop}.
Although these works investigate different data formats, the underlying NPU microarchitectures and computation flows remain similar.
Hence, their protection schemes are easily adaptable across NPUs using different data formats.

\noindent \textbf{Challenges.}
Unlike the conventional formats, BFP redefines both data representation and computation flow, bringing new reliability concerns that existing protection schemes fail to capture: 

\underline{\textit{Where} to protect}:
BFP’s shared-exponent couples elements, meaning a single exponent bit-flip distorts the entire block and propagates amplified correlated errors.
Conversely, extracting the maximum shared-exponent in BFP block gives the mantissa greater significance compared to traditional FP formats, which may cause leading mantissa bit flips to dominate the block’s numerical contribution.
Quantitatively characterizing these error dynamics across hierarchical levels (bit, block, tensor, layer) is non-trivial:
coarse models miss dominant failures, while fine-grained analyses explode in complexity.

\begin{figure}[t]
\captionsetup[subfigure]{skip=2pt} 
    \centering
    \begin{subfigure}[b]{0.46\columnwidth}
        \centering
      \includegraphics[width=\linewidth, height=2.75cm]{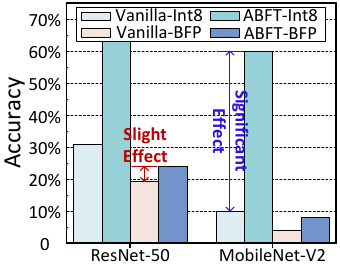}
        \caption{Different models.}
        \label{fig:a}
    \end{subfigure}
    \hfill
    \begin{subfigure}[b]{0.46\columnwidth}
        \centering
\includegraphics[width=\linewidth, height=2.75cm]{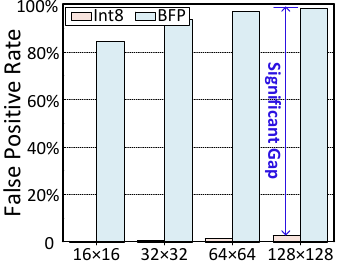}
        \caption{Different configurations.}
        \label{fig:b}
    \end{subfigure}
    \caption{Failure of conventional ABFT-style end-to-end protection on BFP workloads.
    (a) Protection effectiveness of ABFT under INT8 and BFP.
(b) False positive rate of ABFT for INT8 and BFP under different systolic array sizes.
At a fault rate of $5\times 10^{-8}$, the characteristics of BFP introduce intrinsic perturbations in the data, causing the checker’s false positives to skyrocket and rendering the conventional ABFT ineffective.}
    \label{fig:pair}
\end{figure}

\underline{\textit{How} to protect}:
BFP decouples exponent and mantissa processing, demanding a rethinking of fault-tolerance deployment.
As Fig.~\ref{fig:pair} shows, traditional matrix-level schemes (e.g., ABFT) assume linear error accumulation, which fails under BFP's nonlinear block scaling and shared-exponent normalization.
The challenge lies in designing BFP-aware protection that preserves its computational efficiency, selectively protects the most error-sensitive components, and operates within the NPU’s stringent area, latency, and energy budgets.
Striking this balance between protection coverage and performance overhead represents a key open problem toward reliable BFP-based NPUs.

\begin{figure*}[!t] 
\captionsetup[subfigure]{skip=2.5pt} 
    \centering
    \begin{subfigure}{0.32\textwidth}
        \centering
        \includegraphics[width=\linewidth, height=2.2cm]{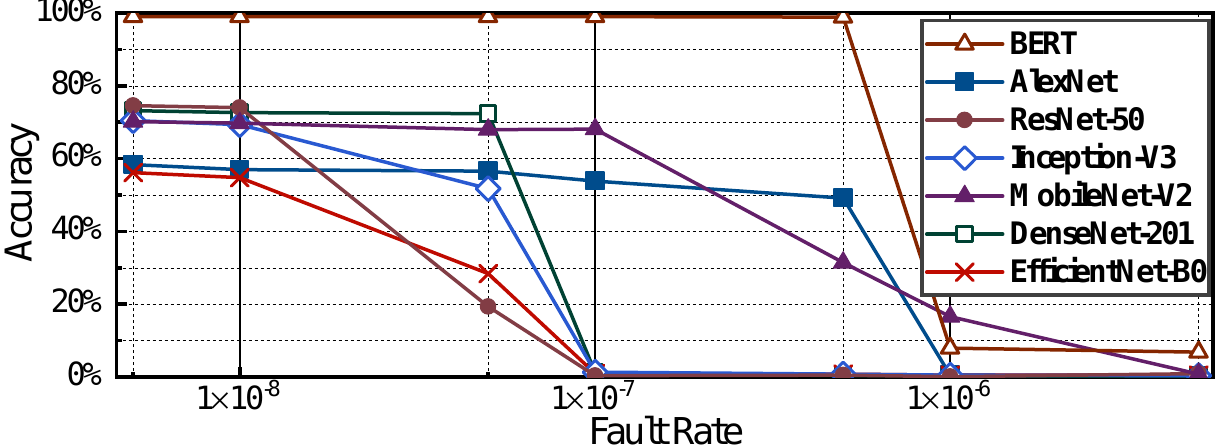}
        \caption{DNNs under different fault rates.}
        \label{dnn_total}
    \end{subfigure}
    \hfill
    \begin{subfigure}{0.32\textwidth}
        \centering
        \includegraphics[width=\linewidth, height=2.2cm]{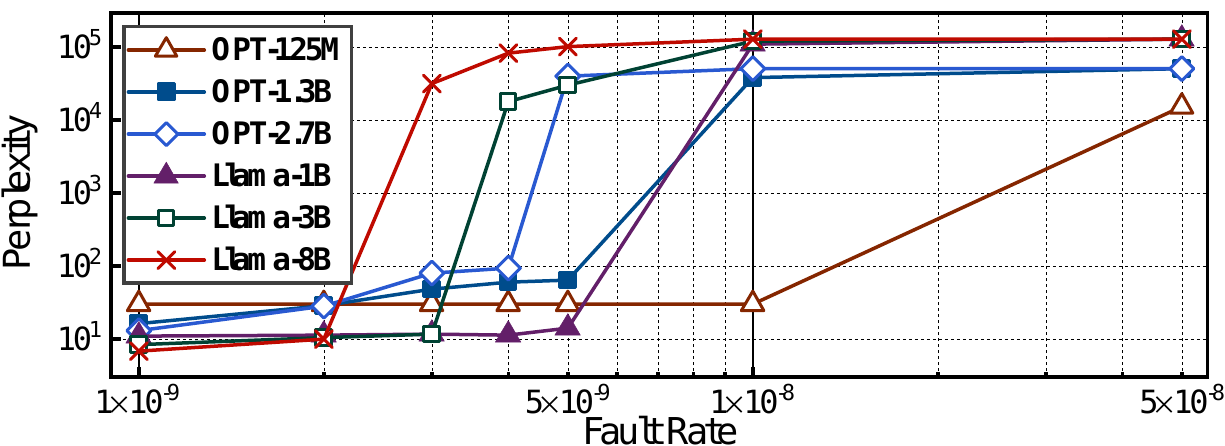}
        \caption{LLMs under different fault rates.}
        \label{llm_total}
    \end{subfigure}
    \hfill
    \begin{subfigure}{0.32\textwidth}
        \centering
        \includegraphics[width=\linewidth]{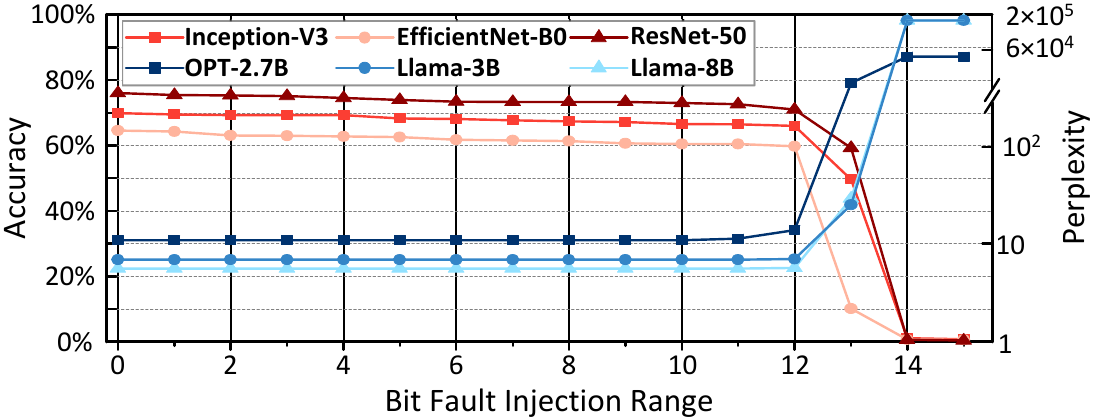}
        \caption{Bit flips at different positions(non-compute modules).}
        \label{bits}
    \end{subfigure}
    
    \begin{subfigure}{0.32\textwidth}
        \centering
        \includegraphics[width=\linewidth]{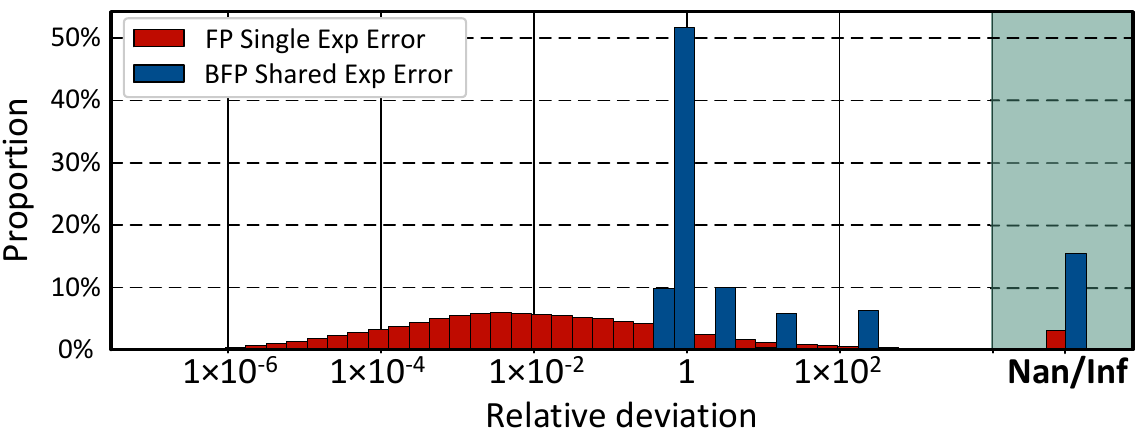}
        \caption{BFP vs. FP in exponent impact(compute modules).}
        \label{exp}
    \end{subfigure}
    \hfill
    \begin{subfigure}{0.32\textwidth}
        \centering
        \includegraphics[width=\linewidth]{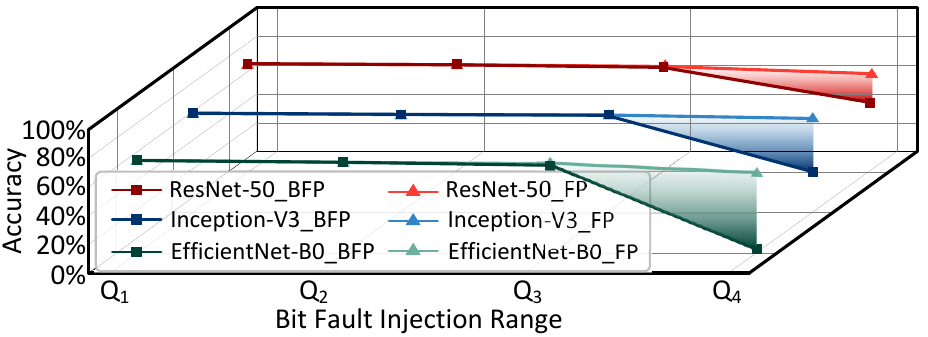}
        \caption{BFP vs. FP in mantissa impact (DNNs).}
        \label{mantissa_dnn}
    \end{subfigure}
    \hfill
    \begin{subfigure}{0.32\textwidth}
        \centering
        \includegraphics[width=\linewidth]{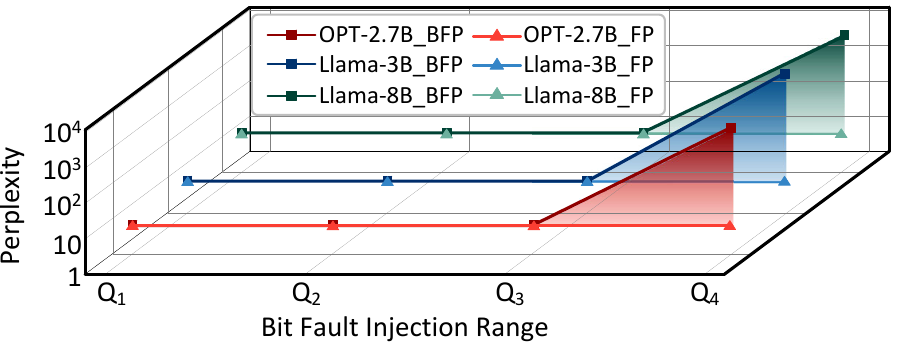}
        \caption{BFP vs. FP in mantissa impact (LLMs).}
        \label{mantissa_llm}
    \end{subfigure}
    \vspace{-5pt}
    \caption{Model accuracy or perplexity under different conditions and comparison of exponent error effects between BFP and FP.
DNNs~\cite{devlin2019bert,huang2017densely,szegedy2016rethinking,he2016deep,sandler2018mobilenetv2,krizhevsky2012imagenet,tan2019efficientnet} were assessed on DBPedia-14~\cite{rangwani2022cost} (BERT) and Tiny ImageNet~\cite{yao2015tiny} (others), and LLMs~\cite{zhang2022opt,dubey2024llama} on WikiText-2~\cite{merity2016pointer}.}
    \label{motivation_0}
\end{figure*}

\noindent \textbf{Contributions.}
We address these challenges via the first in-depth reliability study of BFP-based NPUs.
Using RTL-level fault injection, we characterize failure modes and tolerance thresholds across modules and fault intensities on representative models, revealing key bit-/path-level vulnerabilities: the shared-exponent amplifies correlated errors, and high-order mantissa bits are sensitive due to leading-zero–driven normalization.
Guided by these observations, we propose a BFP-semantic, pipeline-oriented reliability principle and a row/column-wise hardware mapping, decoupling computation into a fixed-point mantissa array and a short scalar exponent pipeline; we exploit exponent-path time slack for low-cost recompute-and-compare, perform fixed-point local consistency checks, and harden compact format-conversion units.
Evaluations show our design achieves near-DMR robustness with 3.55\% average performance (vs. DMR 20\% to 132\%, IR 10\% to 70\%), 1.3\% area and 2.81\% power overheads, $\geq$98\% detection coverage, and sub-microsecond latency.

\section{Preliminary: the Mechanics of BFP}
\label{BFP}

BFP is an efficient data format designed to balance the wide dynamic range with the hardware efficiency of fixed-point.
For a block of $N$ numbers $X = \{x_0, x_1, \dots, x_{N-1}\}$, each element is defined as:
{\scriptsize
\begin{equation}
x_i = (-1)^{s_i} \cdot 2^{e_i} \cdot m_i \quad (0 \le i \le N-1)
\end{equation}
}
The BFP format represents this block by sharing a common exponent, $e_{sh}$.
This shared-exponent is typically chosen as the maximum of all exponents within the block, i.e., $e_{sh} = \max_{0 \le i \le N-1}(e_i)$.
After conversion, the data block can be expressed as:
{\scriptsize
\begin{equation}
\label{eq:bfp_format}
X_{\mathrm{BFP}} = 2^{e_{sh}} \cdot \left\{ (-1)^{s_0}m'_0, (-1)^{s_1}m'_1, \dots, (-1)^{s_{N-1}}m'_{N-1} \right\}
\end{equation}
}
where $m'_i$ is the aligned mantissa, obtained by right-shifting the original mantissa $m_i$ by $(e_{sh} - e_i)$ bits.

The dot product of two $N$-dimensional vectors in BFP format, $\vec{A}$ and $\vec{B}$, with respective shared-exponents $e_A$ and $e_B$, is calculated as:
{\scriptsize
\begin{equation}
\label{eq:bfp_dot_product}
\vec{A} \cdot \vec{B}^T = 2^{e_A + e_B} \sum_{i=0}^{N-1} \left( (-1)^{s_{A,i} \oplus s_{B,i}} \cdot m'_{A,i} \cdot m'_{B,i} \right)
\end{equation}
}
As shown in Eq.~\eqref{eq:bfp_dot_product}, BFP decomposes the dot product into a shared-exponent addition ($e_A + e_B$) and a fixed-point mantissa dot product. Unlike traditional FP, where this separation is confined to the operation-level, BFP elevates it to a macroscopic, block-level. 

\section{Observations: BFP Resilience Characterization}
\label{sc:Obs}


To precisely characterize the behavior of BFP-based machine learning workloads under NPU hardware faults, we conduct systematic fault-injection experiments on a representative set of DNNs~\cite{devlin2019bert,huang2017densely,szegedy2016rethinking,he2016deep,sandler2018mobilenetv2,krizhevsky2012imagenet,tan2019efficientnet} and LLMs~\cite{zhang2022opt,dubey2024llama}. 
By quantitatively analyzing failure modes and end-to-end performance degradation under injection configurations, we delineate the fault-tolerance boundaries of BFP and provide analytical insights into the key questions posed in Sec.~\ref{sc:Introduction}.

\noindent \textbf{Fault-injection setup.} 
Unlike prior studies that perturb per-layer inputs or outputs at the software level~\cite{xie2025realm,reagen2018ares,su2025applying}, abstracting away fine-grained computation and data movement and yielding perturbation patterns misaligned with actual hardware faults, we inject faults at the RTL level to systematically analyze the vulnerability.
We restrict fault sites to SRAM read/write latches and pipeline-boundary registers along the BFP datapath, rather than instrumenting every NPU gate. This follows common error models treating single-event upsets as bit flips in storage elements, disregarding unlatched combinational transients~\cite{geier2020fast,xie2025multidimensional}. We target these locations because they store architecturally visible BFP state and  and determine software-observed results, keeping fault injection complexity tractable.

We model two canonical fault types: permanent faults forcing a bit to stuck-at-0/1 per inference, and transient faults flipping it for one random cycle. Each candidate bit is wrapped by a 2:1-multiplexer “saboteur”~\cite{grinschgl2011automatic,civera2002fpga} that normally forwards the original value, but outputs a forced 0/1 (permanent faults) or the original bit XORed with a one-hot mask (transient faults) during injection. Injection intensity is parameterized by a \emph{fault rate}, defined as injected faults per inference divided by the total number of bits in the system, with locations and timing chosen pseudo-randomly. Because the FPGA prototype cannot host large LLMs, LLM fault-injection experiments use a hardware-semantic software simulator that mirrors RTL fault behavior under the same fault-rate configurations.
\subsection{Analysis of BFP-Based Machine Learning Workloads}

Using the aforementioned fault-injection, we evaluate representative DNNs and LLMs under NPU fault injection.
As shown in Fig.~\ref{motivation_0}(\subref{dnn_total}) and \ref{motivation_0}(\subref{llm_total}), the models maintain baseline performance at sufficiently low fault rates (e.g., below $10^{-9}$); however, as the fault rate increases, DNNs accuracy degrades progressively while LLMs perplexity rises sharply, and when the fault rate exceeds $10^{-8}$ all tested LLMs fail completely. Although  BFP can reduce hardware complexity, it does not provide inherent improvements in fault resilience — once the fault rate surpasses the model’s tolerance threshold, performance deteriorates rapidly. This observation is consistent with prior findings on conventional FP models~\cite{azizimazreah2018soft,bolchini2023resilience}. 
Given that silent-fault rates in modern compute-intensive devices routinely exceed $10^{-7}$~\cite{gizopoulos2025dark}, such intrinsic tolerance is clearly insufficient to ensure accurate operation.

\noindent\emph{\textbf{Insight-1}: BFP-based NPUs require reliability provisioning, as BFP offers no intrinsic resilience and models can catastrophically degrade once fault rates exceed their tolerance thresholds, which are generally below the commonly cited $10^{-7}$ level in industry.}

\subsection{Fault Vulnerability Analysis of BFP-Based NPUs Modules}
We inject faults into distinct BFP-based NPU microarchitectural modules to analyze vulnerability.
Compute and non-compute modules play fundamentally different roles in the dataflow: compute groups (e.g., MAC/ALU pipelines) perform arithmetic, where transient faults quickly amplify via accumulation, whereas non-compute groups (e.g., on-chip memories, buffers) store and transport data, so faults tend to persist and be repeatedly reused. 
These differences create distinct failure modes and naturally call for different protection mechanisms. 
Thus, we separate the analysis into compute and non-compute groups and evaluate them independently, yielding a reliability-oriented abstraction that captures dominant vulnerabilities while avoiding a module-by-module dimensional explosion.

\noindent \textbf{Non-compute modules.} 
In non-compute modules at a high fault rate of $10^{-7}$ (Fig.~\ref{motivation_0}(\subref{bits})), BFP mantissas rarely cause inference failure,
whereas exponent bit upsets are the primary drivers of performance degradation, consistent with observations on conventional FP formats~\cite{li2017understanding,elliott2013quantifying}. 
The rationale is that mantissa perturbations in non-compute units typically manifest as bounded magnitude shifts, while exponent errors rescale values by \(2^n\), producing extreme values. 

\noindent \textbf{Compute modules.} 
Fig.~\ref{motivation_0}(\subref{exp}) shows BFP exponent faults cause much larger deviations than FP due to the shared-exponent design: a single perturbation uniformly rescales all elements in a block, coherently biasing all partial products so accumulation amplifies the error, producing abrupt result spikes, IEEE-754 exceptions (Inf/NaN), and inference failure. For mantissa faults, we analyze the sensitivity of the intermediate-result register along the mantissa path under a fault rate of $5\times10^{-7}$. This register is bit-width expanded to 22 bits for both FP and BFP, and we partition it into four contiguous significance segments: Q1 (the least significant 6 bits), Q2 (the next 6), Q3 (the next 5), and Q4 (the most significant 5). Fig.~\ref{motivation_0}(\subref{mantissa_dnn}) and \ref{motivation_0}(\subref{mantissa_llm}) show BFP mantissa faults degrade performance substantially more than FP, dominated by errors in the highest-significance segment Q4.

\begin{figure}[t] 
    \centering
    \includegraphics[width=\columnwidth,height=3.55cm]{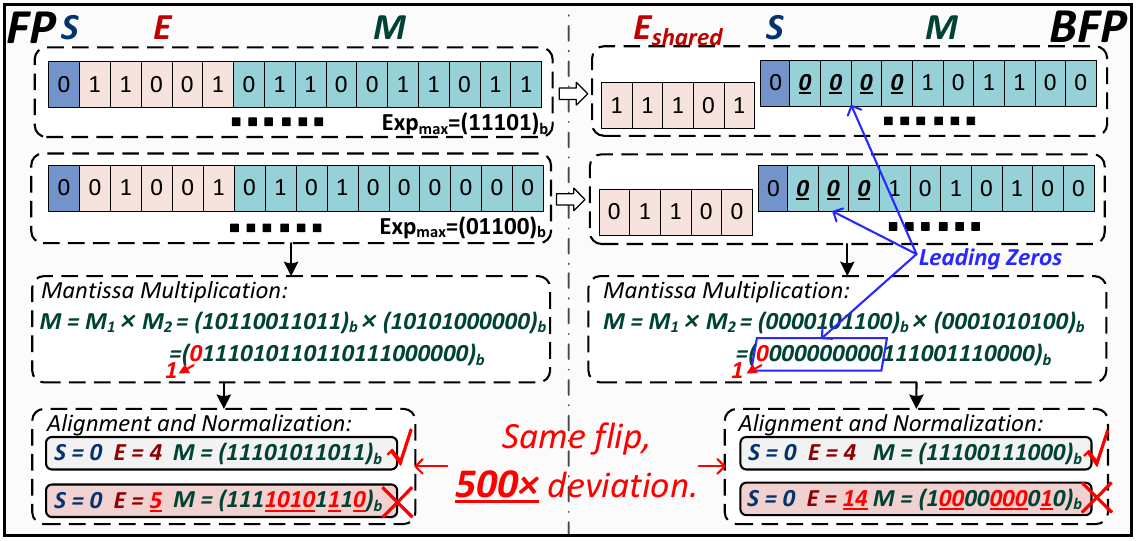} 
    \caption{BFP vs. FP: leading-zero distribution and error amplification. Due to the mantissa right-shift alignment in BFP, intermediate results contain more leading zeros than FP. A bit flip at the same bit position induces a larger deviation in the final normalized value under BFP.}
    \label{motivation_1}
\end{figure}

The discrepancy between BFP and FP stems from their distinct computation pathways. 
BFP follows the computation model in Sec.~\ref{BFP}, with a single post-block normalization subsequently restoring the final exponents and mantissas.
Before the computation, BFP quantization right-shifts mantissas, creating a leading-zero bias. Normalization via a leading-zero counter (LZC) makes this pattern highly sensitive to high-order bit flips. As Fig.~\ref{motivation_1} shows, flipping the first of ten leading zeros collapses the LZC from 10 to 0, increasing the exponent by 10, whereas FP implicit leading bit bounds leading zeros and keeps exponent drift over 500× smaller.
The pipeline ordering also differs: FP performs continuous mantissa alignment and normalization, while BFP first executes block-level mantissa computation and then applies a single, unified normalization.
Consequently, in BFP, normalization allows mantissa errors to perturb output exponents, leading to the assumption that  ``protecting only the exponent path'' is sufficient. Targeted hardening of the mantissa computation path is also required.

\noindent\emph{\textbf{Insight-2}: The bit-level vulnerability of BFP across different modules warrants differentiated hardening priorities.
In non-compute modules, exponent bits should be prioritized.
In compute modules, priority should be given to the shared-exponent path and the high-significance segments of the widened mantissas within execution units.}

\begin{figure}[t] 
    \centering
    \includegraphics[width=\columnwidth,height=7.5cm]{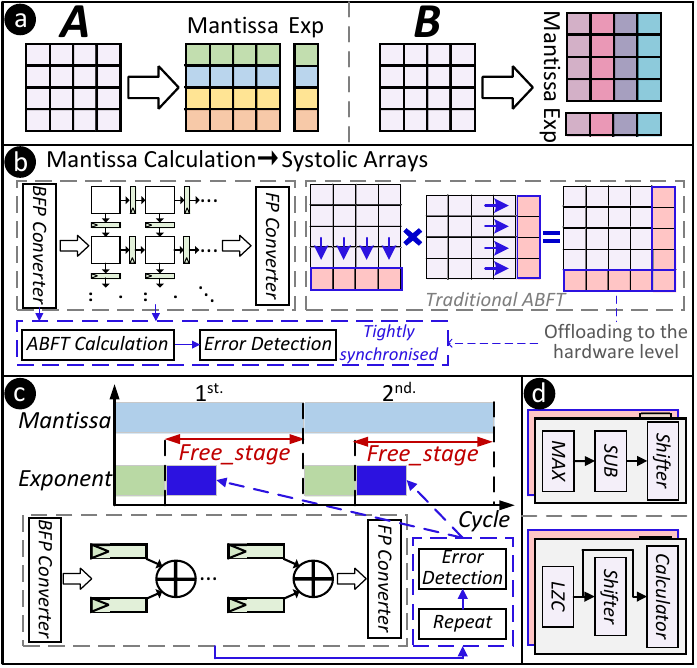} 
    \caption{To balance hardware practicality and reliability, BFP support uses blocking strategy \blacka{} and the mantissa and exponent are mapped to the systolic array and adder respectively. In NPUs, this design modifies or introduces three primary modules: the mantissa compute module (\blackb{}), the exponent compute module (\blackc{}), and the data-format converters (\blackd{}). Each module is provisioned with a dedicated protection mechanism.}
    \label{toplevel}
\end{figure}

\subsection{Analysis of BFP Stream Processing in BFP-Based NPUs}
In typical BFP-based NPU datapath~\cite{fan2019static,zhang2022fast,han2025bbal,drumond2018training}, compute modules are flanked by FP-BFP converters.
This common industrial design confines BFP to inner MACs for efficiency, retaining FP at boundaries for high-accuracy nonlinearities (e.g., softmax) and compatibility with existing FP software stacks.
The pipeline is naturally segmented by these BFP-introduced stages, forming structural boundaries that later serve as anchor points for reliability mechanisms.

Examining BFP's computation logic shows that, when exponents and mantissas are mapped onto decoupled hardware paths, their processing becomes highly asymmetric in structure and verification complexity.
During vector computation, mantissas traverse the MAC pipeline for the dot product, while exponents require only a single scalar addition, leaving the exponent path short with substantial cycle slack. 
Meanwhile, mantissas stay in the fixed-point domain along the dot product datapath, enabling a lightweight fixed-point check without reconstructing FP results.
By contrast, an end-to-end unified check on reconstructed FP outputs must reconcile normalization, alignment, and rounding, increasing checker overhead and false positives that reduce effectiveness (see Fig.~\ref{fig:pair}). 
These observations suggest that aligning BFP hardware mapping with reliability constraints is key to elegant and reliable BFP-based NPUs.

\noindent\emph{\textbf{Insight-3}: By aligning BFP computational semantics with reliability constraints through fine-grained hardware mapping and characteristic-aware protection, a 
cost-effective, reliable BFP-based NPU solution can be developed.}

\section{Microarchitecture}
Based on the analysis in Sec.~\ref{sc:Obs}, we propose a BFP-based NPU that exploits BFP's characteristics for low-overhead reliability.
The design aligns the modified hardware directly with BFP computational semantics and enables highly efficient, minimal-overhead protection mechanisms.
From a reliability perspective,
immediate fault detection is essential to prevent system failures~\cite{iso201126262,TI_FuSa_Manual_2023}.
This priority is codified in ISO 26262~\cite{iso201126262}, which prescribes strict, level-specific requirements for fault detection.
Accordingly, our work focuses on developing efficient fault detection mechanisms.
Once a fault is detected, the system can transition to a safe state and perform corrective actions through well-established safety management frameworks.

\begin{figure}[t] 
    \centering
    \includegraphics[width=\columnwidth, height=3.5cm]{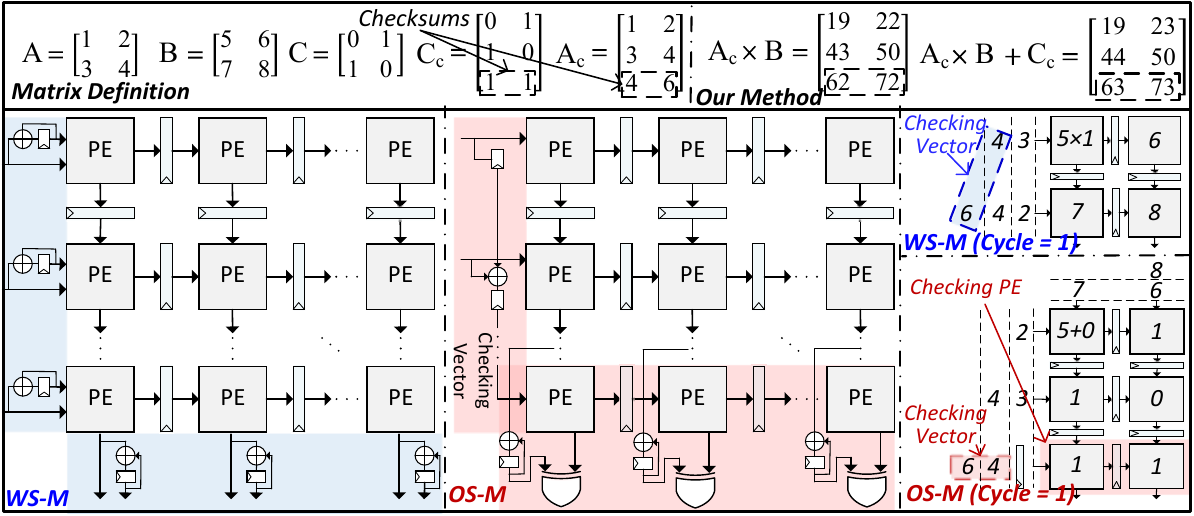} 
    \caption{Mathematical principles (top) and microarchitecture (bottom) for protecting mantissa. Blue regions: protection-driven additions under WS-M; red regions: protection-driven additions under OS-M.}
    \label{systolic_array}
\end{figure}

\noindent \textbf{BFP blocking strategy and hardware mapping.}
Leveraging BFP's characteristics, we design its hardware operators via an appropriate blocking strategy.
BFP's shared-exponent property naturally aligns with vector dot products, making it compatible with systolic arrays.
As shown in Fig.~\ref{toplevel}.\blacka\ , we propose a highly compatible blocking strategy: the left matrix is blocked by rows, the right matrix is blocked by columns, and each vector is quantized into a BFP block with a shared-exponent.
This equates result computation to the dot product of a BFP row and column vector.
Since the final exponent equals the sum of the vectors’ shared-exponents, mantissa MACs are fully decoupled from the exponent path. 
This key decoupling brings several microarchitectural advantages: (i) the mantissa dot product task can be directly mapped to standard fixed-point systolic arrays and vector units, keeping compatibility with traditional NPU hardware without changing key structures; (ii) separating exponent and mantissa pipelines enables independent, differentiated protection schemes exploiting distinct sensitivity characteristics; (iii) mantissa protection reduces to fixed-point protection, avoiding false positives and detection failures from rounding errors in traditional FP paths. Furthermore, once decoupled from the mantissa, the exponent can be obtained via simple summation, requiring only a few adders.

\noindent \textbf{Protection philosophy.}
The above mapping enables comprehensive reliability with minimal overhead.
First, mantissa computation is cast to fixed-point and executed on the systolic array.
Because its significant impact on end-to-end inference and ABFT is deviation-free for fixed-point computations, we deploy a lightweight ABFT checker in hardware to protect the mantissa path (Fig.~\ref{toplevel}.\blackb{}).
In contrast, the exponent computation uses a few adders and is much faster than the mantissa.
Exploiting the timing slack relative to the mantissa pipeline, each adder performs two evaluations within idle cycles and compares the results, providing time-redundant fault detection (Fig.~\ref{toplevel}.\blackc{}).
For FP-to-BFP and BFP-to-FP converters, our measurements show that their area accounts for only 0.2\% of the NPU.
For such small-area, inseparable critical modules,
DMR (Fig.~\ref{toplevel}.\blackd{}) becomes a rational and implementable solution.

\subsection{Reliability of the Mantissa Compute Module}

By mapping BFP mantissa computation onto the systolic array, protecting this path reduces to hardening the fixed-point systolic array.
In NPUs, two common regimes are Weight-Stationary (WS-M) and Output-Stationary (OS-M): in WS-M, the weight matrix  $B$ is preloaded into the systolic array, while the input matrix $A$ streams through; in OS-M, the bias matrix $C$ is preloaded, and both $A$ and $B$ stream in. 
For the fixed-point computation path, we employ
ABFT for verification: by appending row/column checksums, a set of resultant checksums
is generated along with the results (the upper part of Fig.~\ref{systolic_array}).
To keep checks “in-flight” without disturbing the dataflow, we append a checksum vector to the tail of $A$ and, under OS-M, activate additional check-PEs for synchronous in-pipeline verification.

\begin{figure}[t] 
    \centering
    \includegraphics[width=\columnwidth, height=3.5cm]{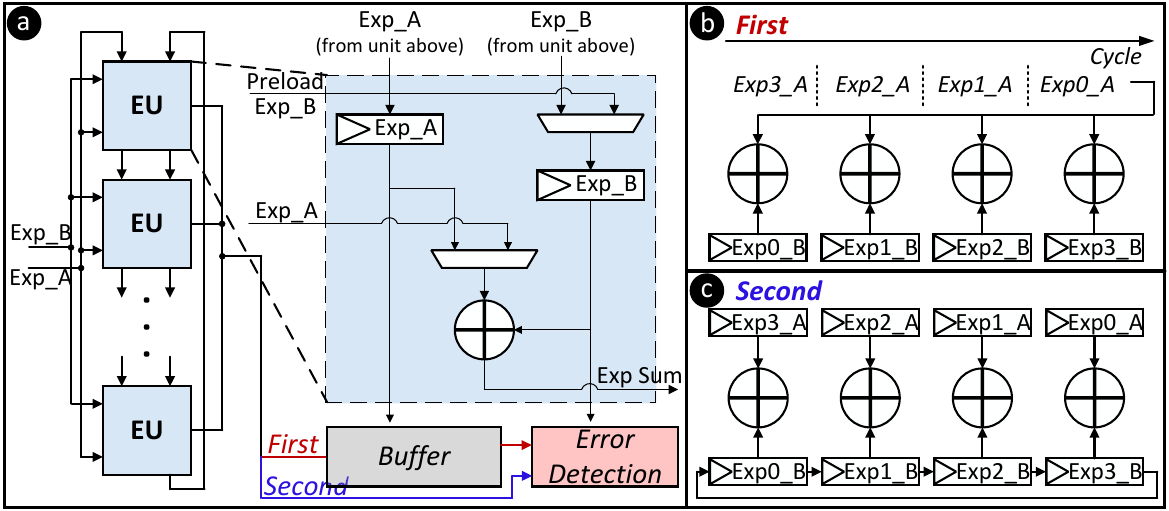} 
    \caption{Microarchitecture of the protection mechanism for the exponent computation module (\blacka{}) and its timing principles (\blackb{}, \blackc{}).}
    \label{Exp}
\end{figure}

\noindent \textbf{Protection mechanism.} 
The mechanism (Fig.~\ref{systolic_array}) tightly integrates with the systolic array’s original dataflow.
In WS-M, we place a column of adder–register units at the array ingress to integrate the streaming columns of the input matrix $A$ on the fly, producing a check-vector that is appended to $A$ and inserted in the same stream. Via systolic propagation, the checksum emerges at the array bottom, 
where an accumulator calculates the sum of the computed results, and a comparator checks the resultant checksums against this sum, 
providing in-situ result verification (Fig.~\ref{systolic_array}: blue region).
In OS-M, outputs accumulate inside the PEs. To support streaming verification, we add (i) a bottom row of check-PEs and (ii) a left-edge column-wise accumulation chain that cascades and integrates per-column inputs each cycle, generating the check components in place. By phase-aligning the chain depth with the array latency, these components arrive at the check-PE inputs exactly when the check-vector is inserted, the check-PEs then produce output-aligned resultant checksums and perform immediate comparison and error detection (Fig.~\ref{systolic_array}: red region).
The mechanism adds only two extra cycles — one to insert the check-vector and one to perform the final comparison. 
Additionally, informed by the Insight-2 in Sec.~\ref{sc:Obs}, the error correction mechanism deployed on our detection system could be designed to ignore bits that exert a negligible influence.

\subsection{Reliability of the Exponent Compute Module}
The exponent compute module for BFP matrix multiplication computes the exponent component of the resultant matrix $C$. 
The module takes two shared-exponent vectors as input: $E_a \in \mathbb{Z}^{N \times 1}$ for the rows of matrix $A$, and $E_b \in \mathbb{Z}^{1 \times N}$ for the columns of matrix $B$. It outputs the resultant exponent matrix $E_c \in \mathbb{Z}^{N \times N}$, where $E_c(i, j)$ is the exponent for the corresponding result element $C(i, j)$. The computation is performed using an outer-add operation, and it is defined by: $E_c = E_a \mathbf{1}^T + \mathbf{1} E_b$. Here, $\mathbf{1}$ is a column vector of ones of size $N \times 1$.
To enhance reliability and optimize resource utilization, a dedicated Exponent Unit (EU) is designed. 
Multiple EUs are cascaded into an EU array that exploits pipeline idle cycles from the timing mismatch between the systolic array and the adder.
The array is capable of executing two aforementioned exponent computations within the time equivalent to a single computation cycle of the systolic array.

\noindent \textbf{Protection mechanism.} 
Fig.~\ref{Exp}.\blacka{}shows the EU microarchitecture and its interconnection topology.
Each EU comprises an adder, two exponent registers and multiplexers.
In WS-M, the systolic array is preloaded with Matrix $B$; in OS-M, Matrix $A$ is fetched and transposed beforehand. 
We next detail the WS-M flow; the OS-M case is analogous to $A$/$B$ swapped and the transpose step applied. 
The exponents of Matrix $B$ are extracted and loaded into a $B$-matrix exponent register in each EU. As Matrix $A$ is fetched rowwise, its row exponents are streamed at one per cycle (Fig.~\ref{Exp}.\blackb{}), broadcast to all EU adders, and latched by a serial $A$-matrix exponent-register chain. 
To increase fault-detection coverage on the second computation and avoid masking permanent faults, the $B$-matrix exponent registers are reconfigured into a ring for data circulation (Fig.~\ref{Exp}.\blackc{}). This causes each adder to receive a different operand pairing across the two runs, reducing constant-input cases that could mask permanent faults.

\begin{figure}[t] 
    \centering
    \includegraphics[width=\columnwidth, height=3.8cm]{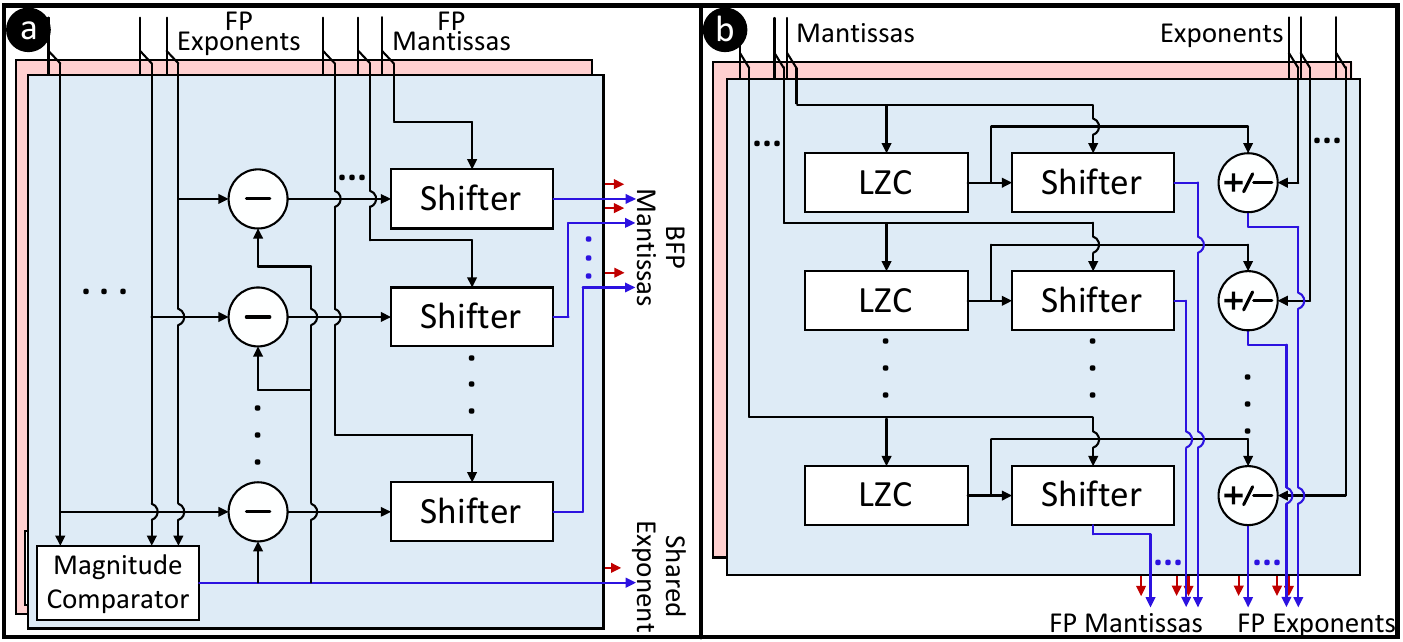} 
    \caption{Protection mechanism and microarchitecture of the FP-to-BFP (\blacka{}) and BFP-to-FP (\blackb{}) converters.}
    \label{converter}
\end{figure}

\subsection{Reliability of the Data Format Converter}

The FP-to-BFP converter operates blockwise to compute a shared-exponent and to align mantissas accordingly. For each block, a magnitude comparator first determines the blockwise maximum exponent. Next, a subtractor derives the relative shift amount for each element. Finally, a shifter right-shifts the corresponding mantissa, and the aligned mantissas are combined with the shared-exponent to form the BFP representation (Fig.~\ref{converter}.\blacka{}).
In the reverse path, the BFP-to-FP converter restores BFP results to the standard FP format. An LZC produces the displacement required to normalize each mantissa. Guided by this displacement, a shifter applies the required adjustment to the mantissa, while an exponent update unit correspondingly updates the output exponent to complete normalization (Fig.~\ref{converter}.\blackb{}).

\noindent \textbf{Protection mechanism.} 
Since Sec.~\ref{sc:Obs} reveals the mantissa's lower criticality on the non-compute path, lightweight fuzzy protection can be applied to mantissa within the converter: in FP-to-BFP converter, FP implicit leading bit fidelity during shifting and the correctness of newly introduced leading zeros after BFP alignment are verified; in BFP-to-FP converter, correctness checks are restricted to a small window of bits immediately following the first significant “1” in the BFP mantissa, which map to the FP mantissa's most significant bits after conversion.
However, standard cell gate-level synthesis shows these converters occupy only 0.2\% of the NPU area.
Thus, even with these beneficial characteristics, the actual gain is minimal due to the already small pre-optimization overhead, which ironically leads to an unfavorable accuracy-reliability trade-off.
Therefore, DMR with bitwise consistency check becomes a direct and rational choice.

\begin{figure}[t]
    \centering
    \includegraphics[width=\columnwidth,height=2.15cm]{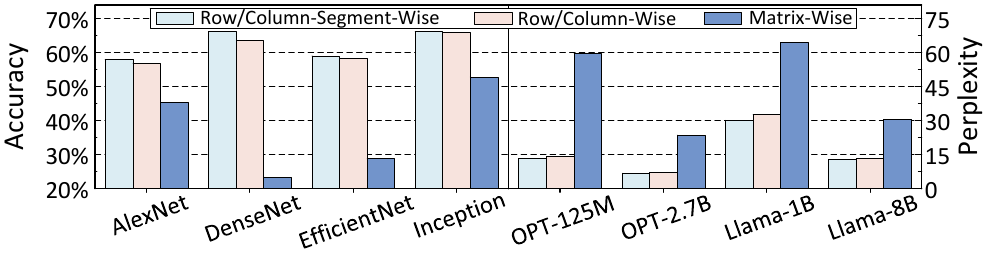} 
    \caption{DNN and LLM performance under three blocking strategies.}
    \label{discussion1}
\end{figure}

\subsection{Discussion on the Blocking Design Trade Off}

We compare the BFP row/column-wise blocking strategy against two common alternatives: matrix-wise~\cite{han2025bbal,koster2017flexpoint} (one shared-exponent for the entire matrix) and row/column-segment-wise~\cite{zhou2025fpga} (each matrix row or column is partitioned into several segments, each with its own shared-exponent), to show why row/column-wise provides a suitable alignment between hardware mapping and reliability design.






\noindent \textbf{Matrix-wise.} 
While a matrix-wise shared-exponent maximizes reuse,
it significantly degrades inference performance (Fig.~\ref{discussion1}). More importantly, it brings negligible practical resource benefit. Relative to the row/column-wise, matrix-wise only reduces exponent computation hardware, and its pipeline latency is much shorter than the mantissa path. 
Thus, although our reliability mechanisms can easily port to this mapping, the cost–benefit profile offers no substantive advantage.

\noindent \textbf{Row/column-segment-wise.} 
Smaller blocks do yield a modest improvement in model performance (see Fig.~\ref{discussion1}), but at the expense of two substantial costs.
(i) Microarchitecturally, it is less friendly: row/column-segment breaks the row-wise scheduling assumption of mainstream systolic NPUs~\cite{genc2021gemmini,jouppi2017datacenter}.
The need to align exponent results from different blocks forces a re-design of datapath and control flow, tightly coupling the hardware to this fine-grained blocking.
(ii) In terms of reliability, the hardware overhead from the fault detection logic grows rapidly.
Since exponents and mantissas are repartitioned at the row/column-segment level, the vulnerability of shared units and intermediate registers also manifests at this granularity: if protection is still deployed at matrix granularity, the detection effect is compromised; if it is instead applied at row/column-segment granularity, then each row/column-segment must be equipped with detection logic, causing overhead to increase sharply as the block size shrinks.
In a $128 \times 128$ matrix, splitting each row/column of four elements raises the hardware overhead from $\approx \frac{1}{128}$ at matrix granularity to $\approx \frac{1}{4}$ at row/column-segment granularity.
Yet, these substantial costs yield only a marginal inference gain ($\approx$2\% in Fig.~\ref{discussion1}).

In contrast, row/column-wise (i) aligns with the row-scheduled model of existing NPUs, allowing reliability mechanisms to be deployed at array granularity to reduce overhead without sacrificing detection accuracy,
and (ii) maintains strong model-level performance.
Thus, it provides a more reasonable convergence point between hardware mapping and reliability design.

\begin{figure}[t]
    \centering
    \includegraphics[width=\columnwidth,height=2.15cm]{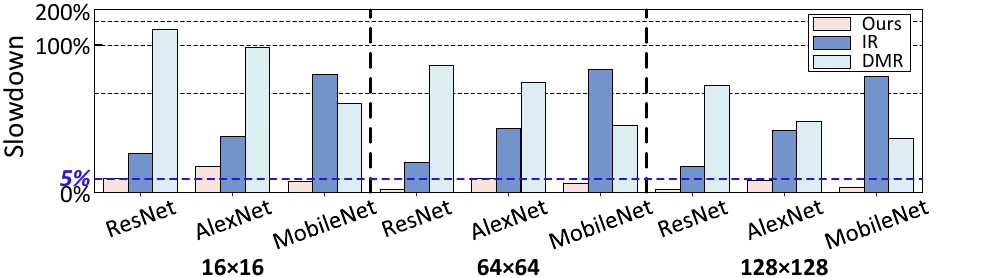} 
    \vspace{-15pt}
    \caption{Performance overhead induced by different strategies.}
    \label{slowdown}
        \vspace{-10pt}
\end{figure}

\begin{figure}[t]
    \centering
    \includegraphics[width=\columnwidth,height=2.15cm]{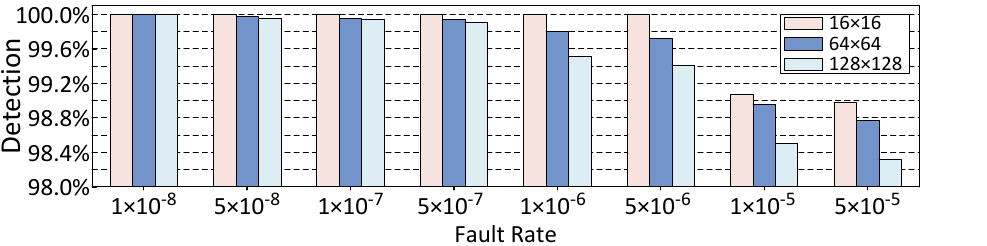} 
        \vspace{-15pt}
    \caption{Error detection rate of our method across different settings.}
    \label{detection1}
        \vspace{-10pt}
\end{figure}

\begin{figure*}[t]
    \centering
    \includegraphics[width=\textwidth]{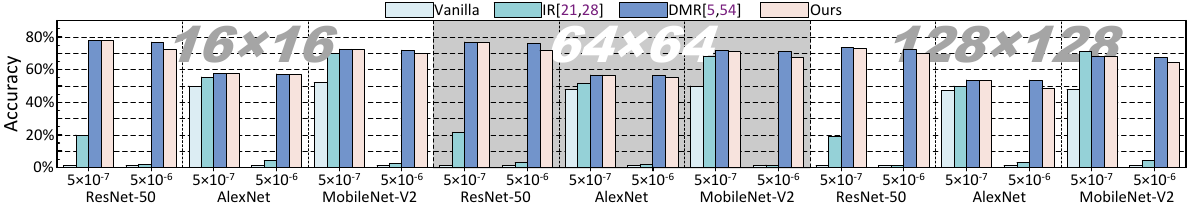} 
    \caption{DNN accuracy under different protection strategies across varying fault rates and hardware configurations.}
    \label{dnn}
\end{figure*}

\begin{figure}[t]
    \centering
    \includegraphics[width=\columnwidth]{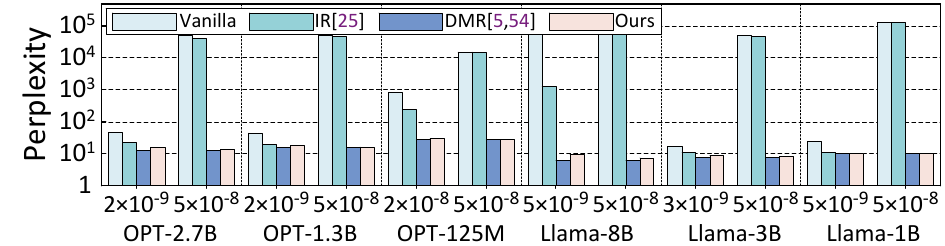} 
    \caption{LLM perplexity under different protection strategies across varying fault rates.}
    \label{llm}
\end{figure}

\section{Evaluation}

\subsection{Experimental Setup}
We implement our design on Gemmini~\cite{genc2021gemmini}, an actively maintained open-source NPU whose systolic array-based core is representative of modern NPU architectures.
RTL is deployed on an FPGA platform comprising 20 AMD Virtex UltraScale+ VU19P devices.
To span a range of scales and resource budgets, we instantiate $16 \times 16$, $64 \times 64$, and $128 \times 128$ systolic arrays. 
Owing to hardware resource limits, LLM performance is assessed in PyTorch using a behavioral model patterned after Spike~\cite{riscv-spike}.
We evaluate mainstream DNNs, including ResNet-50, AlexNet, and MobileNet-V2, on Tiny ImageNet. 
For LLMs, we consider the widely used open-source families Llama3 (1B/3B/8B) and OPT (125M/1.3B/2.7B), assessing them on WikiText-2. Collectively, these choices span a spectrum of architectural styles and parameter scales and, by relying on standard vision and language benchmarks, offer strong cross-study comparability. 
For fault injection, we adopt the same method as in Sec.~\ref{sc:Obs}. 
Then we conduct a comparative analysis against two fault-tolerance baselines that cover representative hardware and software paths: (i) DMR: it is applied to the BFP compute module and the adjacent data format converter; replicas execute in lockstep and their outputs are compared. (ii) IR: a sensitivity-driven dual-execution policy is adopted. Redundancy is prioritized on robustness-critical operators and boundary layers. For DNNs, the protected regions are AlexNet’s first three conv layers, the first two convs in each MobileNet-V2 block, and the first two residual blocks of ResNet-50~\cite{guo2023neural,ibrahim2020soft1}. For LLMs, the earliest 20\% and the final 20\% of layers are protected~\cite{he2025fine}.

\subsection{Observations and Results Analysis}

\noindent \textbf{(1) Performance overhead.}
Fig.~\ref{slowdown} reports the performance overhead of different fault-detection schemes. For a fair comparison, all baselines are evaluated under an iso-area constraint, matched to the proposed hardware implementation. Overall, DMR has the largest overhead (20\%$\sim$132\%) due to extensive hardware replication; moreover, as array size grows and systolic-array utilization drops, DMR’s relative overhead decreases accordingly. IR's overhead is moderate (10\%$\sim$70\%) and correlates strongly with compute density. By contrast, the proposed method adds only minor overhead (geometric mean 3.55\%). 
This indicates the proposed scheme results in substantially lower performance cost, outperforming DMR and IR.

\noindent \textbf{(2) Detection coverage.}
Fig.~\ref{detection1} shows the error detection coverage of the proposed method. Under all configurations and fault rate conditions, the error detection coverage remains above 98\%, with only a slight decrease under extreme fault rate scenarios. This is attributed to the design’s exploitation of the characteristics of BFP computation, which transforms FP-based checks into fixed-point checks.
This approach avoids inconsistencies in results across different computation paths due to rounding errors in FP computations.

\noindent \textbf{(3) Model performance.}
Fig.~\ref{dnn} and ~\ref{llm} present the performance of DNN and LLM models under different configurations and fault rates with various protection schemes. Compared to the Vanilla baseline (no protection), IR provides moderate protection at low fault rates, but its effectiveness quickly decreases at higher fault rates. DMR, due to its comprehensive redundancy approach, consistently exhibits the best protection across different fault rates and configurations. In comparison, the proposed method achieves protection performance comparable to DMR at low fault rates, and second only to DMR at higher fault rates, demonstrating excellent fault tolerance.

\noindent \textbf{(4) Detection latency.}
We evaluate the worst-case error-detection latency, the interval from instruction initiation to error detection.
As shown in Tab.~\ref{detection}, the latency of the exponent and mantissa modules increases with matrix size. Because larger matrix sizes involve more data and complex computations, leading to an increase in latency.
In contrast, the latency for FP-to-BFP and BFP-to-FP converters is shorter and nearly configuration-invariant, as their vector-level pipelines have a fixed number of stages.
Despite the latency increases with the computation scale, all measured latencies remain sub-microsecond, demonstrating the efficiency of the protection scheme.

\noindent \textbf{(5) Hardware overhead.}
We measure the hardware overhead using Synopsys Design Compiler (v2019.12) and the industry-standard TSMC 28nm library.
As shown in Tab.~\ref{Hardware}, in the $128\times128$ configuration, which better aligns with industrial deployments, our scheme incurs only $\approx$3\% power overhead (WS: 2.81\%, OS: 3.05\%) and 1\%$\sim$2\% area overhead (WS: 1.30\%, OS: 1.80\%). These results suggest that the scheme’s linear costs are effectively amortized on larger arrays, thereby yielding a minimal hardware budget for reliability.

\begin{table}[t]
\centering
\caption{The worst case detection latency of various components.}
\vspace{-10pt}
\footnotesize
\label{detection}
\resizebox{0.48\textwidth}{!}{
\begin{tabular}{c c c c c}
\toprule
Configuration &Mantissa compute &Exponent compute &FP-to-BFP &BFP-to-FP \\
\midrule
\textbf{16$\times$16}  &0.098 $\mu$s  &0.064 $\mu$s  &0.014 $\mu$s  &0.008 $\mu$s \\
\textbf{64$\times$64}  &0.390 $\mu$s   &0.250 $\mu$s   &0.018 $\mu$s   &0.008 $\mu$s \\
\textbf{128$\times$128}  &0.750 $\mu$s   &0.524 $\mu$s   &0.020 $\mu$s   &0.008 $\mu$s \\
\bottomrule
\end{tabular}
}
\vspace{-6pt}
\end{table}

\begin{table}[t]
\centering
\caption{Hardware overhead under different configurations.}
\vspace{-10pt}
\footnotesize
\label{Hardware}
\resizebox{0.48\textwidth}{!}{
\begin{tabular}{c c c c c c c}
\toprule
\multirow{2}{*}{Configuration} & \multicolumn{2}{c}{\textbf{$16 \times 16$}} & \multicolumn{2}{c}{\textbf{$64 \times 64$}} & \multicolumn{2}{c}{\textbf{$128 \times 128$}} \\
                              & WS      & OS      & WS      & OS      & WS      & OS      \\
\midrule
\textbf{Overhead (Power)} &22.36\% &23.48\% &4.80\% &6.33\% &\cellcolor{red!25}\textbf{\textit{2.81\%}} &\cellcolor{red!25}\textbf{\textit{3.05\%}} \\
\textbf{Overhead (Area)}  &9.40\% &17.00\% &2.53\% &3.82\% &\cellcolor{red!25}\textbf{\textit{1.30\%}} &\cellcolor{red!25}\textbf{\textit{1.80\%}} \\
\bottomrule
\end{tabular}
}
\end{table}

\section{Conclusion}
In this paper, we present a pathway toward reliable BFP-based NPUs.
We show that effective protection cannot be obtained by directly porting FP/INT schemes but must be co-engineered with BFP’s numerical characteristics and hardware mapping.
Grounded in a systematic RTL-level reliability study, we first answer \emph{where} to protect by identifying dominant module- and bit-level vulnerabilities in different units.
Leveraging these insights, and by further analyzing BFP’s block-wise scaling and stream processing, 
we trace the failure of conventional schemes under shared-exponent normalization and propose a reliability–hardware co-design that realigns blocking, mapping, and protection, thereby answering \emph{how} to protect, ultimately delivering an elegant and reliable BFP-based NPU at minimal costs.

\section{Acknowledgement}
We'd like to thank the reviewers for the helpful feedback. 
This work is supported by the National Key Research and Development Program (Grant No.2024YFB4405600), the Basic Research Program of Jiangsu (Grants No. BK20243042), and the Fundamental Research Funds for the Central Universities (No. 2242025K20013).

\bibliographystyle{ACM-Reference-Format}
\bibliography{ref}

\end{document}